\documentclass[english,aps,preprint]{revtex4}
\usepackage[font={small,it}]{caption}
\usepackage{newtxtext,newtxmath}
\usepackage{mathrsfs}
\usepackage[export]{adjustbox}
\usepackage{physics}
\usepackage[colorlinks=true,citecolor=blue,breaklinks=true,linkcolor=red,
linktocpage=true,urlcolor=blue,pagebackref=false]{hyperref}

\usepackage{natbib}
\usepackage[linesnumbered,ruled,vlined]{algorithm2e}
\SetKwInput{KwInput}{Input}   
\SetKwInput{KwOutput}{Output}   
\SetKwInput{KwFunctions}{Functions}

\begin{document}

	\title{Quantum Semantic Learning by Reverse Annealing an Adiabatic Quantum Computer}
	
	\author{Lorenzo Rocutto}
    \thanks{Current affiliation of Lorenzo Rocutto: Università degli Studi di Bologna}
	\affiliation{Istituto di Fotonica e Nanotecnologie, Consiglio Nazionale delle Ricerche, Piazza Leonardo da Vinci 32, I--20133 Milano, Italy}
	\affiliation{Dipartimento di Fisica, Universit\'a degli Studi di Milano Bicocca, Piazza della Scienza, 3--20126 Milano, Italy}
	\author{Claudio Destri}
	\affiliation{Dipartimento di Fisica, Universit\'a degli Studi di Milano Bicocca, Piazza della Scienza, 3--20126 Milano, Italy}
	\author{Enrico Prati}
    \thanks{Corresponding author: Enrico Prati - Email: enrico.prati@cnr.it}
	\affiliation{Istituto di Fotonica e Nanotecnologie, Consiglio Nazionale delle Ricerche, Piazza Leonardo da Vinci 32, I--20133 Milano, Italy}
	\begin{abstract}

Boltzmann Machines constitute a class of neural networks with applications to image reconstruction, pattern classification and unsupervised learning in general. Their most common variants, called Restricted Boltzmann Machines (RBMs) exhibit a good trade--off between computability on existing silicon--based hardware and generality of possible applications. 

Still, despite several successful achievements, the diffusion of RBMs is quite limited, since their training process proves to be hard. The advent of commercial Adiabatic Quantum Computers (AQCs) raised the expectation that the implementations of RBMs on such quantum devices could increase the training speed with respect to conventional hardware.

To date, however, the implementation of RBM networks on AQCs has been limited by the low qubit connectivity when each qubit acts as a node of the neural network.

Here we demonstrate the feasibility of a complete RBM on AQCs, thanks to an embedding that associates its nodes to virtual qubits, thus outperforming previous implementations based on incomplete graphs.

Moreover, to accelerate the learning, we implement a semantic quantum search which, contrary to previous proposals, takes the input data as initial boundary conditions to start each learning step of the RBM, thanks to a reverse annealing schedule.

Such an approach, unlike the more conventional forward annealing schedule, allows sampling configurations in a meaningful neighborhood of the training data, mimicking the behavior of the classical Gibbs sampling algorithm.

We show that the learning based on reverse annealing quickly raises the sampling probability of a meaningful subset of the set of the configurations. As a consequence, even without a proper optimization of the annealing schedule, the RBM semantically trained by reverse annealing achieves better scores on reconstruction tasks.

More generally, our work paves the way towards the establishment of a quantum advantage of Adiabatic Quantum Computers, especially given the foreseen increased connectivity and number of qubits of the next generations of quantum hardware.

    \end{abstract}
    \maketitle
    \section*{Introduction}

Unsupervised learning algorithms are expected to be empowered by the advent of practical quantum computing. Nonetheless, the achievement of a computational advantage over the Von Neumann--Zuse computer architecture \cite{schmidhuber2006colossus} based on classical binary logics for useful--sized problems is still on its way, mainly since quantum hardware is in an early stage of development. To speed up the process, hardware synthesis techniques \cite{mueck2017quantum} must be constantly updated to harness the full potential of the quantum processing units.

Boltzmann Machines (BMs) are a prominent example of unsupervised learning algorithm \cite{barlow1989unsupervised}, \cite{le2013building}, introduced in 1985 by Ackley, Hinton and Sejnowski \cite{ackley1985learning}. It has been proved that a trained BM can be used as a universal approximator of probability distributions on binary variables \cite{sussmann1988learning},\cite{younes1996synchronous}. Such property makes BMs particularly suitable if used as a generative model \cite{theis2015note}, \cite{srivastava2012multimodal} to reconstruct partially missing data. Despite such theoretical representative power, one of the steps of the training algorithm makes computationally expensive to train large BMs \cite{hinton2002training}. More specifically, such a step requires to extract samples from the instantaneous distribution approximated by the BM. The cost to produce each sample grows rapidly as the problem size increases. Thus, BMs are usually not applicable in useful--sized problems.

To achieve the full power of BMs, a training method that scales well as the size and the complexity of a BM increase would be required. A possible solution follows from the observation that BMs can be naturally implemented on a specific architecture of quantum computers referred to as Adiabatic Quantum Computers (AQC). Being able to train useful--sized BMs on an AQC would enable to tackle and solve relevant problems in many fields ranging from recommendation systems \cite{bell2007lessons}, to anomaly detection \cite{yang2017improved},\cite{rastatter2019abnormal}, to quantum tomography \cite{carrasquilla2019reconstructing} .

During the years, BMs have found application mostly in a simplified form, called Restricted Boltzmann Machine (RBM) \cite{smolensky1986information}, \cite{hinton2002training}. RBMs have lower representative power than BMs, but they are easier to train and use. Due to their popularity in the classical version, RBMs are usually chosen as a benchmark to evaluate the potential advantages of the quantum learning algorithms \cite{benedetti2016estimation}. 

Since 2011, researchers explored the possibility to use an AQC to approximate the probability distribution needed during the training of an RBM \cite{denil2011toward}, \cite{dumoulin2014challenges}. Experimental results suggest that an AQC can be operated to extract samples from the distribution associated with an RBM, at the computational cost of a single quantum operation  \cite{amin2015searching}, \cite{boixo2016computational}, \cite{korenkevych2016benchmarking}. Thus, sampling from AQCs seems to have a cost that does not depend on the size or the complexity of the RBM. In 2015, Adachi and Henderson  \cite{adachi2015application} performed computations on an actual AQC to train a Deep Belief Network, a directed graph version of RBM. In 2016, Benedetti et al. \cite{benedetti2016estimation} introduced an efficient technique for estimating the effective temperature of the distribution produced by the AQC. Developing hardware synthesis techniques to sample effectively from a given probability distribution is beneficial for unsupervised learning in general. For instance, a hybrid algorithm has been recently proposed where a classical Generative Adversarial Network (GAN) is supported by a quantum--trained RBM \cite{anschuetz2019near}.
Previously, some of the authors exploited deep learning algorithms \cite{prati2017quantum} to achieve control of qubits for preparation of quantum states \cite{porotti2019coherent}, \cite{porotti2019reinforcement}, \cite{paparelle2020digitally}. Here we follow the opposite approach since we combine machine learning with the computational power of the quantum hardware itself.  

Following the practice of benchmarking BMs through a restricted version, we tackled the problem of training an RBM using an AQC \cite{biamonte2017quantum}.  Results obtained in such a restricted case are fundamental building blocks for more complex networks.

We applied embedding techniques to prepare a graph of virtual qubits arranged to implement all the weights present in a classical RBM. We compare our results to those obtained with a network characterized by sparse connections between the visible and the hidden layers \cite{benedetti2016estimation} to show the improvement in the learning.
Differently from circuital quantum computers based on quantum logic gates \cite{rieffel2011quantum}, adiabatic quantum computers \cite{childs2001robustness,boixo2016computational,johnson2011quantum,farhi2002quantum,dickson2013thermally} provide the quantum analog of simulated annealing so they return probabilistic answers by driving a quantum system from a superposition of classical states to a frozen state which approximates the solution of a problem. 
Such process is called quantum annealing and it may implemented on adiabatic quantum computers. Unfortunately, the commonly adopted forward annealing schedule fails to entirely use the information carried by the elements of the dataset to start the search as it starts from a generic multi--qubit state.   
Contrary to previous algorithms, we introduce semantic learning based on the reverse annealing schedule, which consists of partially relaxing the multi--qubit state from a chosen initial classical state, performing a local quantum search in the configurations space.

The results obtained by applying the reverse annealing schedule take advantage of the ability to perform the annealing process starting from configurations belonging to the dataset. For this reason it is referred to as semantuc and it is more informative if compared to the forward annealing method.

To assess the performance of the training process we trained an RBM with 16 visible units and 16 hidden units to reconstruct the Bars and Stripes dataset. The performance is evaluated in terms of the percentage of reconstructed pixels and the average Log--Likelihood of the dataset. All the quantum computations have been performed on a D--Wave 2000Q\textsuperscript{TM} System processor \cite{johnson2011quantum}, \cite{dwave2019technical}, \cite{dwave2017reverse}.

We show that embedded topologies do not sensibly interfere with the capacity of the AQC to produce correctly distributed samples. The use of embedding techniques thus allows faster learning if compared to the sparse case.

The sampling distribution produced by the AQC in the case of the reverse annealing schedule is closer to the configurations corresponding to the dataset than that produced by the more conventional forward annealing. As a consequence, the learning process exhibits a novel semantic behavior that brings pros and cons. In our case, the RBM trained with the use of reverse annealing achieves overall a better reconstruction score if compared to the usual quantum algorithm for AQCs. Furthermore, the use of reverse annealing leads to a sampling probability of elements of the dataset which is double that of the forward annealing and higher than that of the classical method. 

In the Results Section (\ref{sec:results}), the classical RBM is briefly introduced and the quantum training method for the RBM is described, both in its forward annealing and reverse annealing schedule. The experimental results for both approaches are shown. In Section \ref{sec:discussion} the results are discussed. The Method Section (\ref{sec:Methods}) describes the technical details and the design of the experiment. Finally, Section \ref{sec:conclusions} summarizes the main conclusions and the perspectives.

\section{Results}
\label{sec:results}

\subsection{Structure of BM and RBM}

\begin{figure}[]
    \centering
  \includegraphics[width=0.9\textwidth]{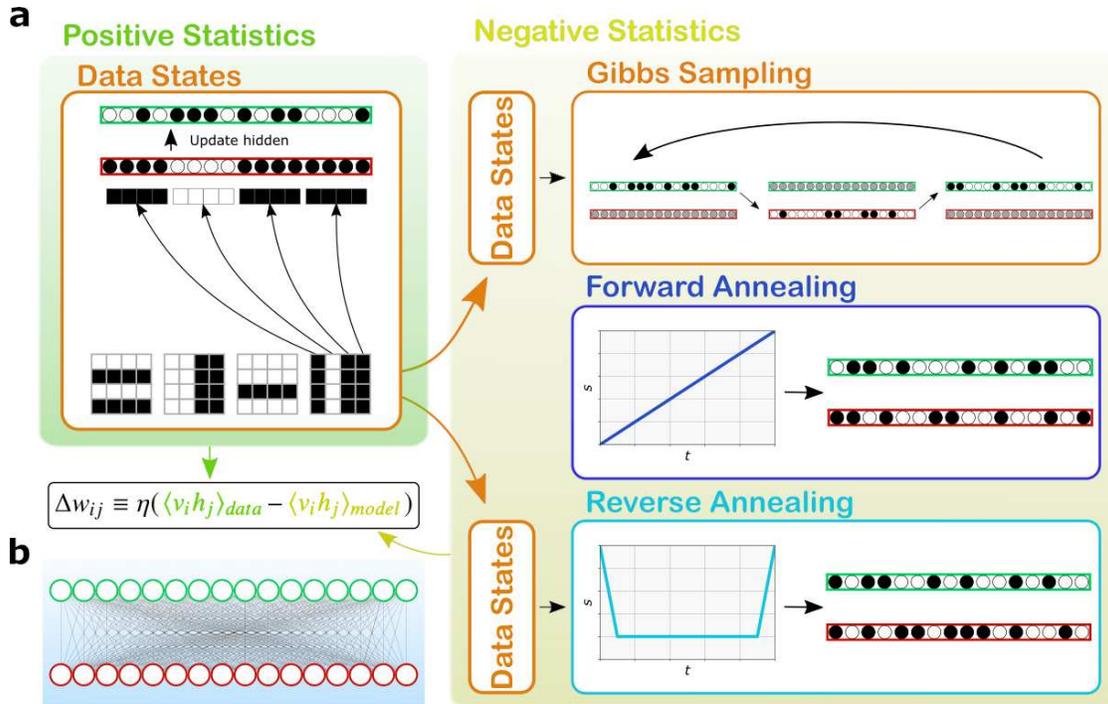}
  \caption{\textbf{a}: Schematic overview of the methods to train a Restricted Boltzmann Machine. The green and yellow boxes show which part of the gradient is estimated by each step of the algorithm. Steps that require only classical computations are circled in orange. During the data states phase, the elements of the dataset are loaded in the visible units (red), and the hidden units (green) are updated accordingly. The result allows the estimation of the positive statistics. Next, one step among Gibbs, forward annealing, and reverse annealing sampling is chosen to estimate the negative statistics. Forward annealing is the only one that does not depend on states loaded from the dataset. \textbf{b}: Structure of the Restricted Boltzmann Machine used in the article, with 16 visible units fully connected to 16 hidden units.}
  \label{fig:how_to_train_RBM}
\end{figure}

A Boltzmann Machine is represented by a graph composed of units that can assume binary values $\{0,1\}$, linked by real weighted connections \cite{ackley1985learning}. Units are divided into two classes, \emph{visible} and \emph{hidden}. In general, a BM has connections between any pair of units while the graph of an RBM is bipartite, with no hidden--hidden or visible--visible connections. Figure \ref{fig:how_to_train_RBM} \textbf{b} shows the structure of a 16x16 RBM. Before the training, the weights are initialized randomly, according to some probability distribution. A training process that suitably modifies the weights of the BM is needed to encode information. A trained BM can accept an incomplete item and reconstruct the missing parts according to the previously acquired information.
In the case of image reconstruction, both known and missing pixels of the image are univocally represented by a visible unit. The visible units correspond to both the input and the output of the generative model. Indeed, when reconstructing a corrupted image, we initialize and fix part of the visible units, while the uninitialized visible units output the missing pixels. Hidden units, on the other hand, confer to the BM the power to extract essential features from the dataset. A higher number of hidden units raise the ability of the BM to mimic the structure of the dataset. 

\subsubsection{Internal dynamics of a BM}

In 1985, Ackley, Hinton, and Sejnowski \cite{ackley1985learning} proposed a computational algorithm that produces configurations distributed according to an energy model summarized in Supplementary Note \textbf{1}.

First, visible units are initialized with arbitrary binary values. Next, the values of the hidden units are updated according to a probabilistic rule that depends on the state of visible units. Then, visible units are updated according to the state of hidden units. Such an update is usually iterated many times, and the output configuration is achieved by reading the final values of the units. If the number of update steps is sufficiently high, the output configurations are extracted from a distribution which approximates the probability from the energy model 
\begin{equation}
P(S) \equiv \frac{e^{-\frac{E(S)}{T}}}{Z(T)}\;,
\label{eq:prob_model}
\end{equation}
where $Z(T) \equiv \sum_{\{S\}} e^{-E(S)/T}$, $E(S)$ is a dimensionless function depending on the state $S$ of the BM units that we will call \emph{energy}, and $T$ is a dimensionless parameter called \emph{temperature}, in analogy with thermodynamic functions of statistical physics. In the case of an RBM,

    \begin{equation}
    \label{eq:defEforRBM}
        E\big(S\big) = -\sum_{i=1}^{n_v}\sum_{j=1}^{n_h}w_{ij}v_ih_j
        -\sum_{i=1}^{n_v}a_iv_i-\sum_{j=1}^{n_h}b_jh_j
        \;.
    \end{equation}
    
where $v_i$ is the value of the $i$-th visible unit, $h_j$ is the value of the $i$-th hidden unit, $w_{ij}$ is the weight of the connection between the $i$-th visible unit and the $j$-th hidden unit, while $a_i$ and $b_j$ are real parameters called \emph{biases}. Since different values of the temperature $T$ correspond to different uniform rescalings of the weights and biases, we are allowed to set $T=1$. 

The sampling algorithm is described in detail in Supplementary Note \textbf{2}. Such a method to extract configurations is well established in Physics, and it is usually referred to as \emph{Gibbs sampling}. We call $n_g$ the chosen number of times visible units are updated.

\subsubsection{How to train a BM}

\begin{figure}[h]
  \begin{center}
  \captionsetup{width=0.9\textwidth}
  \includegraphics[width=0.8\textwidth]{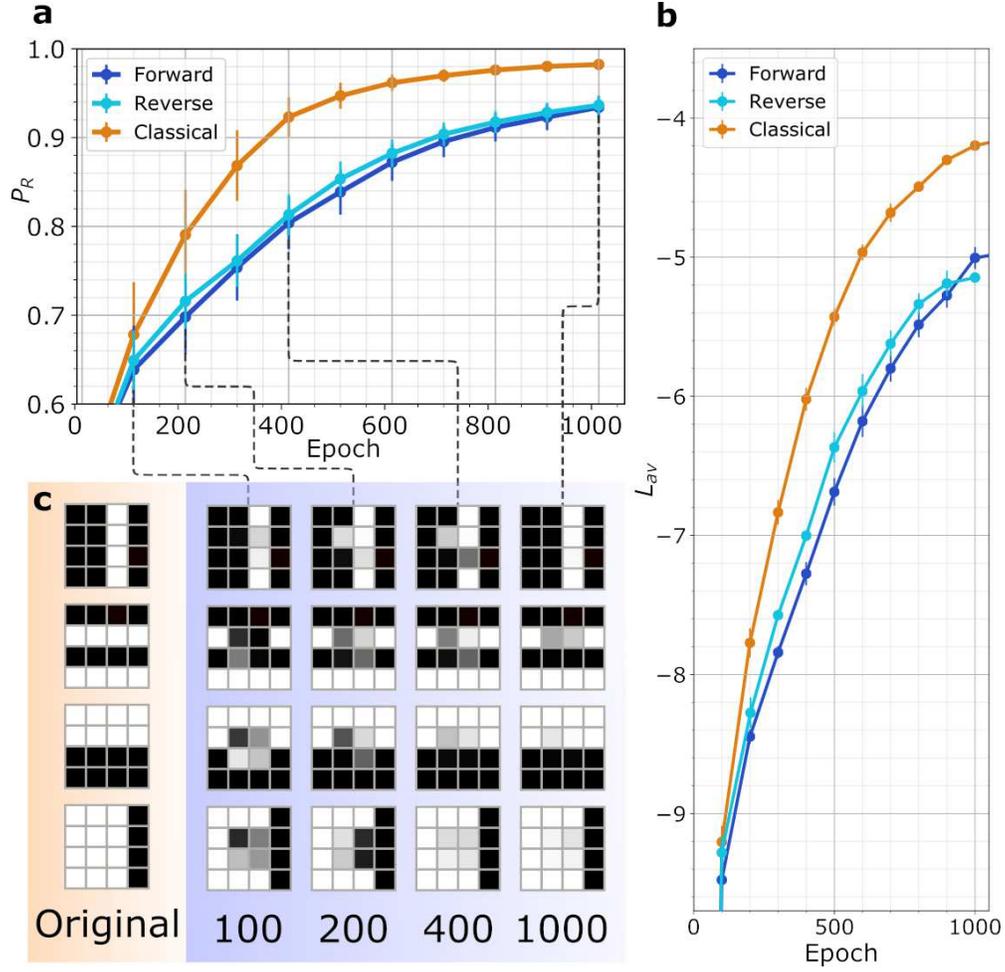}
  \caption{\label{fig:results} \textbf{a}, Reconstruction score vs training epochs for the classical, forward annealing and reverse annealing case, respectively. The learning curves  are obtained with a rescaling parameter $\alpha=0.32$ and the weights initialization parameters  $\mu=0$, $\sigma=2$ and $I_W=[-3,3]$. \textbf{b}, Average Log--Likelihood vs training epochs for the classical, forward annealing and reverse annealing case. \textbf{c}, Some of the images used during training, taken from the $4x4$ Bars and Stripes dataset, together with the reconstruction obtained with $n_g=500$ steps of Gibbs sampling at different epochs.}
  \end{center}
\end{figure}

In the following, the training of the BM involves a dataset of images composed of $M\times M$ pixels, known as Bars and Stripes dataset (BAS). The images consist of either black or white stripes or columns. Figure \ref{fig:results} \textbf{c} shows some sample images taken from the $4\times 4$ version of such dataset. 

Each training image can be mapped to a $M^2$--dimensional binary vector $\mathbf{r}$, where each term in the vector corresponds to a visible unit of the BM. The probability for the BM to output a training vector $\mathbf{r}$ can be raised by adjusting the weights and biases to lower the energy of that vector and to raise the energy of other vectors.
    
The learning rule for performing stochastic steepest ascent in the log probability of the training data is expressed by:
    
    \begin{equation}
    \label{eq:wupdate_steepascent}
        \Delta w_{ij}\equiv \eta (\langle v_i h_j\rangle_{data} - \langle v_i h_j\rangle_{model})\;,
    \end{equation}
where $\eta>0$ is the \emph{learning rate} \cite{ackley1985learning}.

The term $\langle v_i h_j\rangle_{data}$ is called \emph{positive statistics}. It represents the expected value of $v_i h_j$ when the visible units are fixed to the values of a training vector, averaged over all vectors of the training dataset. It is defined as

\begin{equation}
\label{eq:exp_data}
    \langle v_i h_j\rangle_{data}\equiv\frac{1}{N_\mathcal{D}}\sum_{\mathbf{r}\in \mathcal{D}}\left(\frac{\sum_{\{\mathbf{h}\}} r_i h_j e^{-E(\{h\},\mathbf{r})}}{\sum_{\{\mathbf{h}\}} e^{-E(\{h\},\mathbf{r})}}\right)\;,
\end{equation}
where $\mathcal{D}$ is the training set, and $N_\mathcal{D}$ is the number of elements in $\mathcal{D}$.
    
Instead, $\langle v_i h_j\rangle_{model}$ is called \emph{negative statistics} and represents the expectation value of $v_i h_j$ when both visible and hidden units are produced by the model (e.g. during the Gibbs sampling algorithm). It is defined as
\begin{equation}
\label{eq:exp_model}
    \langle v_i h_j\rangle_{model}\equiv\frac{\sum_{\{\mathbf{v}\},\{\mathbf{h}\}} v_i h_j e^{-E(\{\mathbf{v}\},\{\mathbf{h}\})}}{\sum_{\{\mathbf{v}\},\{\mathbf{h}\}} e^{-E(\{\mathbf{v}\},\{\mathbf{h}\})}}
\end{equation}

In the RBM case, since there are no direct connections between hidden units, it is fairly easy to get an unbiased sample of $\langle v_i h_j \rangle_{data}$. A single Gibbs sampling step to update the hidden units is necessary, repeated for each $\mathbf{r}\in\mathcal{D}$ (see Supplementary Note \textbf{1}).

Getting an unbiased sample of $\langle v_i h_j \rangle_{model}$ is much more difficult. From Equation \ref{eq:exp_model} it follows that an exact computation would involve a summation over all possible combinations of values for hidden and visible units, which means summing over $2^{N}$ elements. In many practical situations, an exact computation of the negative gradient is unfeasible.

The negative statistics is usually approximated as follows. Initially, the values of the visible units are set equal to those of a training vector. Next, the Gibbs sampling algorithm is repeatedly applied. More precisely, a sequence of configurations is produced by alternatively extracting the values of the hidden units at fixed visible units and the values of visible units at fixed hidden units, while averaging the value of $v_i h_j$ along the sequence. Each step of this sequence has the same computational cost of $\langle v_i h_j \rangle_{data}$. This process is repeated for each element of the dataset. This provides an estimate of $\langle v_i h_j \rangle_{model}$, which is indeed unbiased if a good level of convergence, a.k.a. \emph{equilibration}, is attained for each sequence.  
RBMs typically learn better if more steps of alternating Gibbs sampling are used before collecting vectors for the negative statistics. Initializing the visible units with elements from the dataset is not a mathematical requirement since average values at equilibrium ought to be independent of the initial conditions, but it is known to enhance the quality of the learning \cite{hinton2002training}.

The Gibbs sampling algorithm allowed to train RBMs in some cases (e.g. the Netflix prize challenge \cite{amatriain2013big}). On the contrary, unrestricted BMs are still out of reach even with this importance--sampling algorithm, due to the presence of hidden--hidden connections which make each step of the alternating Gibbs sampling much more difficult. Indeed, nowadays there is hardly any known application of BMs in real--sized problems.  In principle our findings can be extended to BMs as well.

From now on we focus on the RBM case. Indeed, even if fully connected BMs could gain a greater advantage from the quantum training process, RBMs are usually chosen as a benchmark due to their popularity in the classical framework. 

Note that also biases need to be trained, but their update requires no additional overhead if we can effectively estimate the negative part of the statistics. Such property is explicitly discussed in Supplementary Note \textbf{1}.

\subsection{Quantum--trained RBM}
\label{sec:qrbm}
 
Training an RBM is made easier by a device capable of producing the configurations needed to estimate $\langle v_i h_j \rangle_\text{model}$ within a few operations. It has been suggested that Adiabatic Quantum Computers (AQCs) produced by D--Wave Systems can extract binary configurations distributed as the expression of $P(S)$ in Equation \ref{eq:prob_model} \cite{denil2011toward}.

AQCs are quantum processors that tackles optimization problems by driving a quantum system from a superposition of classical states to a frozen state corresponding to an approximate solution \cite{boixo2016computational}, \cite{johnson2011quantum}, \cite{farhi2002quantum}, \cite{dickson2013thermally}. 

The qubits composing the processor are forced to evolve according to an Ising Hamiltonian that encodes the problem to solve. The total Hamiltonian implemented on D--Wave devices takes the expression  \cite{dwave2019technical}, \cite{morita2008mathematical}, \cite{hajek1988cooling}:
    \begin{equation}
    \label{eq:isingmodelannealing}
    \begin{split}
        H(t)&=-F(s(t))\left(\sum_{i,j}J_{ij}\sigma^z_i\sigma^z_j+\sum_iA_i\sigma^z_i\right)-G(s(t))\sum_i\sigma^x_i\\ &\equiv F(s(t))H_P+G(s(t))H_T\;,
    \end{split}
    \end{equation}
where $t$ is the physical time and $\sigma_i^z$, $\sigma_i^x$ are the Pauli matrices that act along the $z$ and $x$ direction, $J_{ij}$ and $A_i$ are real parameters chosen by the user. The annealing process ideally returns the classical spin configuration that minimizes $H_P$. 
\begin{figure}[]
    \centering
  \captionsetup{width=1\textwidth}
  \includegraphics[width=0.5\textwidth]{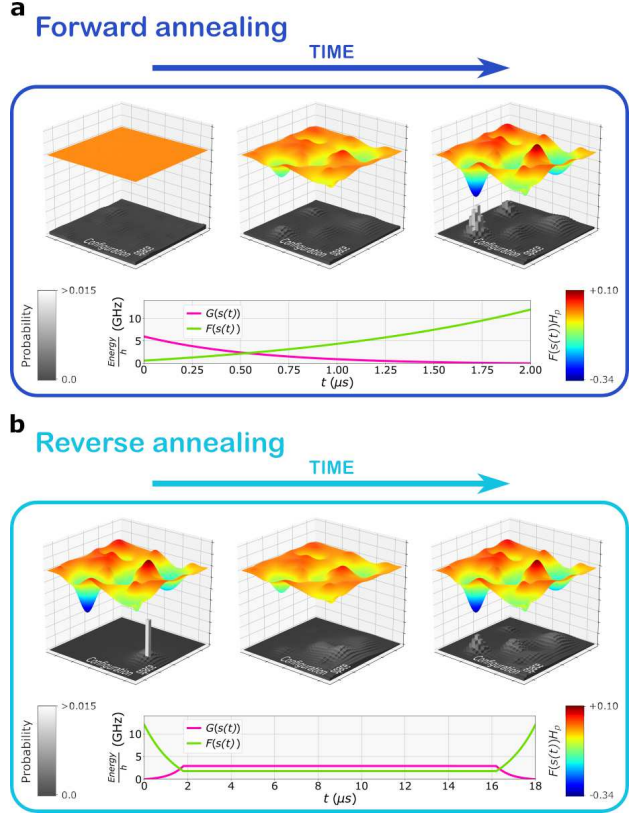}
  \caption{Visual representation of the process according to both quantum forward and reverse annealing schedule. Each bin of the gray 2--D histograms represents a possible configuration produced by the AQC. The height of a bin represents the probability that the corresponding configuration is the output of a single annealing process. The colored surface is a representation of the problem Hamiltonian $H_P$ in Equation \ref{eq:HP}. Each configuration corresponds to a different value of $H_P$. For clarity, the problem is reduced to only two dimensions, and neighboring configurations differ by a single bit. A further graphical simplification consists of smoothing the surface making it continuous. \textbf{a}, During the \textit{forward annealing}, the $\sigma^z$ component of the annealing Hamiltonian increases from $0$ to its maximum value. The behavior is represented by the increase in the weight of $H_P$. The configurations histogram, initially almost flat, becomes more and more peaked around the configurations that minimize $H_P$. \textbf{b}, During the \textit{reverse annealing}, the $\sigma^z$ component of the annealing Hamiltonian $H(t)$ (Equation \ref{eq:isingmodelannealing}) is initially set at its maximum value and the $\sigma^x$ component is null. It is then possible to initialize the system such that a single configuration has probability almost equal to $1$ (apart from noise effects). The annealing is then performed in reverse, lowering $\sigma^z$ components and raising $\sigma^x$ components in the annealing Hamiltonian $H(t)$. The system partially relaxes, and configurations close to the initial one become populated according to the corresponding value for $H_P$, thanks to quantum tunneling and thermal effects. The final step consists of standard forward annealing.}
  \label{fig:figura_uno}
\end{figure}

The function $s(t)\in[0,1]$ is called \emph{annealing schedule}. During the usual annealing schedule, $s=t/t_\text{f}$, where $t_\text{f}$ is the total time of the annealing process, and thus $s$ increases linearly from $0$ to $1$. Such a schedule is usually called \emph{forward annealing}, and its shape is represented in Figure \ref{fig:how_to_train_RBM} \textbf{a}. Figure \ref{fig:figura_uno} \textbf{a} shows how $G(s(t))$ and $F(s(t))$ vary with respect to the time. In the case of the D--Wave 2000Q\textsuperscript{TM} System processor, the user can specify few distinct points during the annealing where $s$ can be set to any value in $[0,1]$ \cite{dwave2017reverse}. In particular, the schedule is called \emph{reverse annealing} if $s=1$ at the start of the annealing. The value $s=1$ corresponds to $G(s(t))=0$, which means the $\sigma^x$ component of the qubits is not coupled to  any transverse field. As a consequence, each qubit can be put in a classical state, $-1$ or $+1$, because spin flip is forbidden \cite{dwave2017reverse}. Next, the parameter $s>0$ is reduced, therefore allowing spin flipping. Figure \ref{fig:how_to_train_RBM} \textbf{a} presents the shape of a reverse annealing schedule. Figure \ref{fig:figura_uno} \textbf{b} shows how $G(s(t)$ and $F(s(t))$ vary with respect to the time in this case.

Previous works have exploited reverse annealing to boost non--negative matrix factorization \cite{ottaviani2018low}, to enhance genetic algorithms \cite{king2019quantum}, to solve portfolio optimization problems \cite{venturelli2019reverse}.

In general, mapping a problem on an AQC and running an annealing cycle returns a configuration that is not the ground state of $H_P$. The reason resides in the sources of noise affecting the hardware \cite{dumoulin2014challenges}, \cite{dwave2019technical},  which determine the violation of the hypothesis of the adiabatic theorem. Dumoulin et al. \cite{dumoulin2014challenges} observe that the erroneous, higher--energy configurations produced by the AQC are approximately distributed as if they were extracted from a Boltzmann distribution at an effective dimensionless temperature $T_\text{eff}$:
    \begin{equation}
    \label{eq:boltzhypothesis_Teff}
        P(\mathbf{v},\mathbf{h}) = \frac{e^{-\frac{E_P(\mathbf{v},\mathbf{h})}{T_\text{eff}}}}{\sum_{\{\mathbf{v}'\},\{\mathbf{h}'\}}e^{-\frac{E_P(\mathbf{v}',\mathbf{h}')}{T_\text{eff}}}}\;,
    \end{equation}
where $E_p(\mathbf{v},\mathbf{h})$ is the eigenvalue corresponding to the eigenstate $(\mathbf{v},\mathbf{h})$ of Hamiltonian $H_P$ (see also \cite{amin2015searching}, \cite{raymond2016global}). The effective temperature $T_\text{eff}$ is a real parameter determined by the presence of thermal noise inside the chip \cite{dwave2019technical}. Its time evolution is considered difficult to forecast \cite{adachi2015application}, \cite{benedetti2016estimation}. The key idea consists of exploiting such a property to produce configurations that are distributed appropriately for the RBM training.

\subsubsection{Mapping the RBM on an adiabatic quantum computer}

We now show how to set the parameters of a D--Wave AQC such that the configurations produced are distributed according to Equation \ref{eq:boltzhypothesis_Teff} at an effective temperature $T_\text{eff}=1$.

First, we introduce two sets of qubits, named $\nu$ and $\lambda$. The set $\nu$ contains $N_v$ qubits, each one corresponding to a visible unit. $\lambda$ contains $N_h$ qubits corresponding to the hidden units. $H_P$ is rewritten as
    \begin{equation}
    \label{eq:HP}
        H_P=\sum_{ij\in\rho}J_{ij}\sigma^z_i\sigma^z_j+\sum_{i\in\nu}A_i\sigma^z_i+\sum_{j\in\lambda}B_j\sigma^z_{j}
    \end{equation}
where $A_i$, $B_j$, $J_{ij}$ are real parameters. The summation in the first term is to be performed on the set $\rho$ of couplers linking visible and hidden qubits.

Note that the RBM energy functional in Equation \ref{eq:defEforRBM} now appears very close in shape to $H_P$ in Equation \ref{eq:HP}. Indeed, the two expressions can be mapped into each other if a suitable rescaling of the parameters is performed. In particular, note that the units of the RBM assume binary values in $\{0,1\}$, while the eigenvalues of $\sigma^z$ belong to $\{-1,1\}$. We can recover the same energy levels in the two expressions by the following mapping:

\begin{equation}
\label{eq:mapping_parameters_RBM_to_AQC}
\begin{aligned}
    J_{ij} &= \alpha\cdot \left(\frac{w_{ij}}{4}\right)\\
    A_i &= \alpha\cdot \left(\frac{a_i}{2}+\sum_{j\in\lambda}\frac{w_{ij}}{2}\right)\\
    B_j &= \alpha\cdot \left(\frac{b_j}{2}+\sum_{i\in\nu}\frac{w_{ij}}{2}\right)\;,
\end{aligned}
\end{equation}
where $\alpha$ is an experimental estimate of the parameter $T_\text{eff}$ that appears in Equation \ref{eq:boltzhypothesis_Teff}. If the AQC parameters are initialized as in Equation \ref{eq:mapping_parameters_RBM_to_AQC}, each annealing cycle samples a probability distribution similar to that appearing in Equation \ref{eq:boltzhypothesis_Teff} \cite{adachi2015application}, \cite{benedetti2016estimation}. In other words, the AQC produces a configuration needed for the training at the cost of a single quantum operation.
    
\subsection{Estimation of the negative statistics by forward annealing}
\label{sec:Estimation of the negative statistics using an AQC}

The estimation of $\langle v_i h_j\rangle_{data}$ is fairly easy on a classical processor, therefore we look at the quantum advantage when estimating $\langle v_i h_j\rangle_{model}$.

The customary quantum algorithm for the estimation of $\langle v_i h_j\rangle_{model}$ \cite{adachi2015application}, \cite{benedetti2016estimation}, assumes that the value of $T_\text{eff}$ can be considered constant over time. Such assumption is tricky since it has been proved that a wrong guess for $T_\text{eff}$ can lead to severe errors during training \cite{benedetti2016estimation}. Nonetheless, a reasonable guess allows to effectively train an RBM \cite{benedetti2016estimation},\cite{benedetti2017quantum}, \cite{henderson2019leveraging}. 

We trained an RBM composed of 16 visible units and 16 hidden units. We trained it on the Bars and Stripes (BAS) dataset defined in \cite{mackay2003information}. The dataset contains images composed of $N\times N$ black or white pixels. In each image, the pixels are arranged in monochromatic bars or stripes. Since our RBM has 16 visible units, we used $4\times 4$ images. Figure \ref{fig:how_to_train_RBM} \textbf{b} shows the structure of the chosen RBM.

We trained five times the RBM for $1100$ epochs using quantum forward annealing. Each instance was initialized with random weights and biases extracted from the same probability distribution. The weight initialization spread $\sigma$ is the value of the standard deviation of the Gaussian used for randomly generate the weights and the biases. The weight initialization range $I_W$ defines the symmetric range centered in zero truncating the Gaussian from which the weights and the biases are sampled. The amplitude of the interval $I_W$ is bound below by the S/N ratio due to both the uncertainty in the values of the currents and the random fluctuations of the chip. It is also bound above by the maximum values of the current. Among the amplitudes tested, $I_W=[-3,3]$ was sufficiently big to overcome weights and sufficiently small to not reach the maximum values allowed for the currents.
Supplementary Figure 1 presents the learning curve for an RBM initialized with smaller weights. The result shows that, if the aforementioned boundaries are respected, the initial distribution of the weights could have a reduced impact on the learning speed.

For each run we set $\alpha=0.32$. Figure \ref{fig:results} \textbf{b} shows the results obtained by averaging over the five runs, in terms of the average Log--Likelihood of the dataset $L_\text{av}$, defined as follows:

\begin{equation}
   L_\text{av}=\frac{1}{N_\mathcal{D}}\sum_{\mathbf{v}\in\mathcal{D}}\log\left(\sum_{\{\mathbf{h}\}}P(\mathbf{v},\mathbf{h})\right)\;,
    \label{eq:loglikelihood}
\end{equation}
where $P(\mathbf{v},\mathbf{h})$ is calculated as it appears in Equation \ref{eq:boltzhypothesis_Teff}, with the hypothesis that $T_\text{eff}=1$. The most expensive step in computing $L_\text{av}$ is obtaining the explicit expression of the partition function $Z(T_\text{eff})$ of the distribution $P(\mathbf{v},\mathbf{h})$ (i.e. the denominator in Equation \ref{eq:boltzhypothesis_Teff}). The number of units in the RBM (32 total) is sufficiently low so that we could calculate $L_\text{av}$ without resorting to approximations.

We also trained five times the RBM for $1100$ epochs using the classical algorithm. The instances were initialized with the weights already used in the forward annealing case. It means that for each quantum instance, there is a classical instance that used the same weight initialization.

\subsection{Estimation of the negative statistics by reverse annealing}
\label{sec:reverse annealing}

To introduce a semantic search emulating the initialization of the Gibbs sampling of the classical case on a quantum adiabatic computer, we adopt the reverse annealing schedule. To our knowledge, such an initialization method has never been used to train an RBM on an AQC. Indeed, the customary quantum annealing method allows a unique system initialization, corresponding to the equally--probable quantum superposition of all the classical states. On the contrary, reverse annealing allows us to perform a quantum search in the neighborhood of the configuration set by the user as the initial state. It then constitutes a possible way to emulate the initialization of the Gibbs sampling in the classical case. 

We modified the usual quantum algorithm for the estimation of $\langle v_i h_j\rangle_{model}$, which exploited forward annealing, to exploit a sampling procedure based on reverse annealing (see Supplementary Note \textbf{2} for the original algorithm).

    \vspace{8 pt}
\begin{algorithm}[H]
\label{alg:deonly}
\DontPrintSemicolon
  \vspace{8 pt}
  \KwInput{Number of visible units $n_v$, number of hidden units $n_h$, desired number of iterations $N_{\text{iter}}$, initial weights $w_{ij}$, initial biases $a_i$ and $b_i$, annealing schedule $s(t)$ such that $s(0)=s(1)=1,\; 0<s(t)<1\;\forall t\neq0,1$, training dataset $\mathcal{D}$ composed by $N_\mathcal{D}$ elements. We suppose $T_{\text{eff}}^{(j)}$ is known for each step $j$.}
  \KwOutput{Matrix $ans_{pq}$ of dimensions $(n_v,n_h)$ that contains the estimates for $\langle v_p h_q\rangle_{model}\;\forall p,\,\forall q$.}
  \KwFunctions{$eval(T_{\text{eff}}, w_{ij}, a_i, b_j)$ returns values for $J_{ij}$, $A_i$ and $B_j$ according to Eq. \ref{eq:mapping_parameters_RBM_to_AQC}; $initialize\_qubits(\mathcal{D}, k)$ initialize each visible units of the RBM with the corresponding pixel in the $k$-th element of $\mathcal{D}$, while hidden units are updated at $T=1$ with a single step of Gibbs sampling, then qubits of the AQC are initialized with the corresponding values mapped in $\{-1,+1\}$; $anneal(J_{ij}, A_i, B_j,s(t))$ uses quantum annealing with $H_P(J_{ij}, A_i, B_j)$ as final Hamiltonian (Eq. \ref{eq:HP}) and $s(t)$ as annealing schedule; $measure(x_i)$ retrieves the binary value of $x_i$ after the annealing process ends.}
  
  $ans_{pq} \leftarrow 0\;\;\forall p,\;\;\forall q$\;
  \For{$m=0 \mathbf{\;to\; } N_\mathcal{D}$}
   {
        \For{$k=0 \mathbf{\;to\; } N_{iter}/N_\mathcal{D}$}
        {
           initialize\_qubits$(\mathcal{D},m)$\;
           $[J_{ij}, A_i, B_j]\leftarrow$ eval$(T_{\text{eff}}^{(j)}, w_{ij}, a_i, b_i)$\;
           anneal$(J_{ij}, A_i, B_j)$\;
           $v_p \leftarrow$ measure$(v_p) \;\;\forall p$\;
           $h_q \leftarrow$ measure$(h_q) \;\;\forall q$\;
           $ans_{pq} \leftarrow ans_{pq}+ v_ph_q\;\;\forall p,\;\;\forall q$
        }
   }
   $ ans_{pq} \leftarrow \frac{ans_{pq}}{N_{iter}}\;\;\forall p,\;\;\forall q$
\caption{Quantum algorithm for the negative statistics estimation using reverse annealing}
\end{algorithm}
\vspace{15 pt}

If compared to the usual procedure, the $N_\text{iter}$ annealing cycles are performed separated in $N_\mathcal{D}$ groups. For each group, qubits corresponding to visible units are initialized with an element from the dataset. Qubits corresponding to hidden units are initialized with binary values classically computed with a single step of Gibbs sampling. After all elements of the dataset have been used to initialize the annealing process, the output configurations are used to estimate $\langle v_i h_j\rangle_{model}$.

Figure \ref{fig:how_to_train_RBM} \textbf{a} reports a schematic overview of the different methods presented in the article that can be used to train an RBM.

We trained twice the RBM for $1000$ epochs using reverse annealing. The instances were initialized with two weights set randomly chosen from the five already used in the classical and forward annealing case. We chose $\alpha=0.32$ as in the forward annealing case.

Figure \ref{fig:results} \textbf{b} shows the results in terms of the average Log--Likelihood of the dataset, compared with the results obtained in the forward annealing case.

\section{Discussion}
\label{sec:discussion}

As one might expect, thanks to the full connectivity based on the embedding on virtual qubits, our forward annealing approach obtains Log--Likelihood scores sensibly better than sparse implementations \cite{dumoulin2014challenges,benedetti2016estimation}. To compare, contrary from Ref. \cite{benedetti2016estimation} where a $16\times16$ RBM is implemented with $80$ visible to hidden connections, our implementation realizes all $256$ connections. At the epoch $1000$ our forward annealing method obtains a score of $-5.00 \pm 0.08$ which is better than the best forward annealing-based result\cite{benedetti2016estimation}, optimized step-by-step by a temperature estimation tool, which reaches a score of $~-5.3$ at iteration $5000$. Even without using any $T_\text{eff}$ estimation technique, the higher number of connections was sufficient to gain a great advantage over the sparse implementation. For the sake of completeness, the same sparse embedding of Ref. \cite{benedetti2016estimation} has been tested at a fixed temperature (see Supplementary Figure 2). As pointed out in Ref. \cite{dumoulin2014challenges}, the complexity of the topology of the AQC seems to be a major bottleneck for the implementation of RBMs.
\begin{figure}[]
  \begin{center}
  \captionsetup{width=0.9\textwidth}
  \includegraphics[width=1\textwidth]{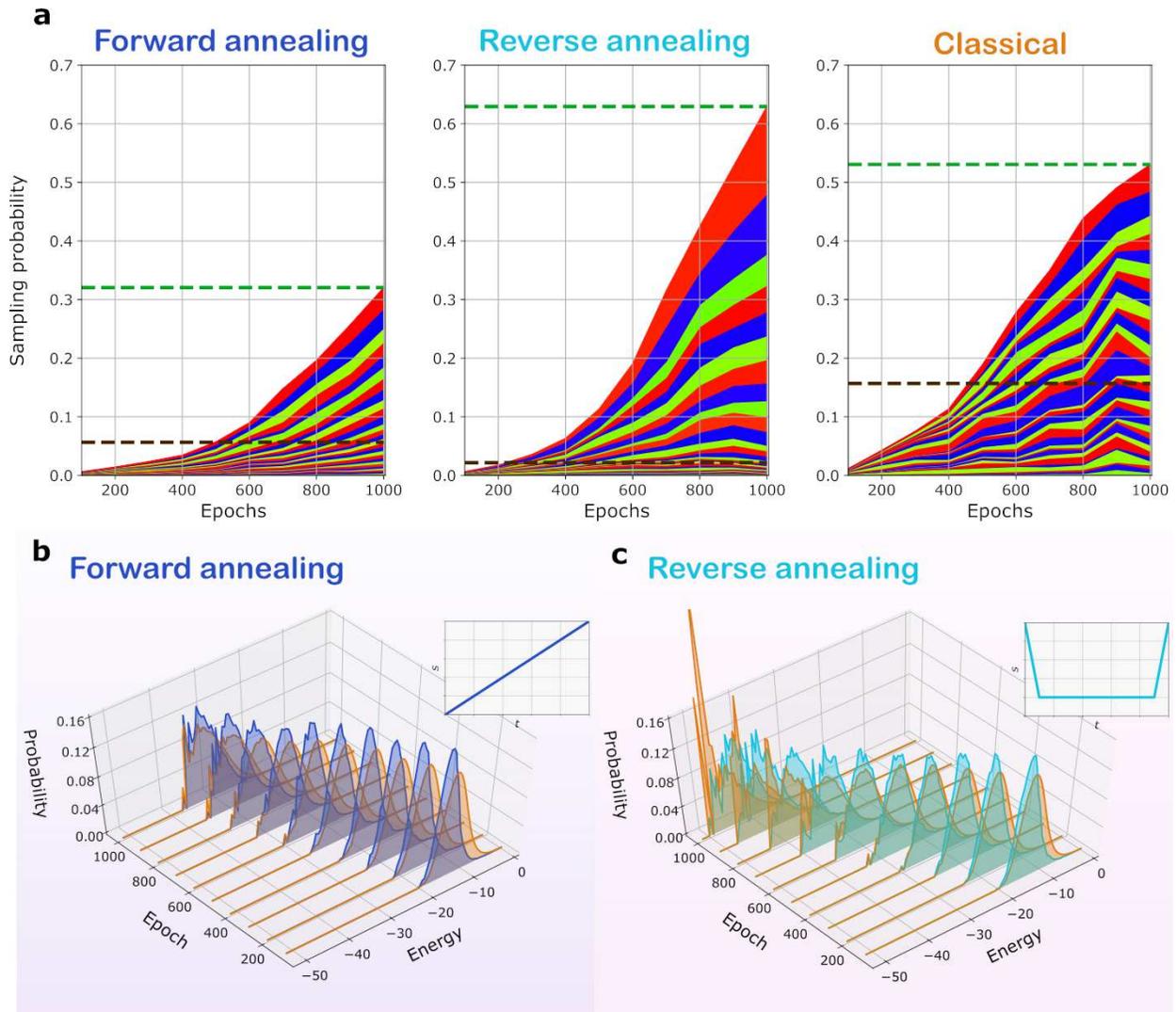}
  \caption{\label{fig:bands} The distributions presented are reconstructed from the weights obtained during the training processes presented in Figure \ref{fig:results}. \textbf{a}, Each colored band represents the probability that the corresponding image of the dataset is sampled from the Boltzmann distribution defined by the RBM at that epoch. Thus, each plot contains $30$ bands. The green line represents the probability that any of the images is sampled at epoch $1000$. The black line represents the probability that any of the $15$ less probable images is sampled at epoch $1000$. \textbf{b} and \textbf{c} represent the distributions produced by the D--Wave machine using forward annealing (\textbf{b}, in blue) and reverse annealing (\textbf{c}, in light blue) each $100$ epochs, from epoch $100$ to epoch $1000$, starting from the same weight initialization used in \textbf{a}. In both \textbf{b} and \textbf{c}, the  Boltzmann distribution at $T=1$ corresponding to the instantaneous value of the RBM weights is represented in orange. The qualitative shape of the annealing schedule $s(t)$ is also shown as reference.}
  \end{center}
\end{figure}    

The value $\alpha=0.32$ was chosen for the forward annealing case because it achieves the better Log--Likelihood score at epoch $1000$, among a range of tested values. We kept the poor man's choice of $\alpha=0.32$ also in the reverse annealing case, and we set the shape of $s(t)$ without further exploration of the hyperparameter space of the annealing schedule. Despite such lack of fine optimization, the reverse annealing performs even better than the forward annealing up to epoch $900$, therefore proving capable to speed up the learning in the initial phase, and comparable at the epoch $1000$, if we limit the evaluation to the Log-Likelihood.\\
We turn now our attention to the differences among energy distributions during the learning by applying the two paradigms. Both the forward and reverse annealing approaches initially produce a distribution corresponding to $T<1$ (Figure \ref{fig:bands} \textbf{b} and \ref{fig:bands} \textbf{c}), since the quantum--sampled distributions are visibly cooler than the theoretical distribution calculated at $T=1$. Such behavior is a consequence of the choice to keep $\alpha$ fixed. The value of $\alpha$ has been chosen to maximize the score at iteration $1000$ for the forward annealing case, so the same value is not necessarily the best during the first iterations.

At epoch $1000$, the lowest energy configurations for the forward case are at energy $\sim(-38)$, while in the reverse case the lowest achieved energy at the same epoch is $\sim(-50)$, where the values are dimensionless, as obtained by the definition of $E(s)$ (Equation \ref{eq:defEforRBM}). Despite the two cases share the same learning rate $\eta$, the training based on reverse annealing modifies the RBM weights so that some configurations correspond to considerably lower energies. We will now explore the consequences of such behavior.

Let's now consider the configurations where the visible units correspond exactly to one of the elements of the dataset. We define $\Delta$ the set composed of such configurations.
$\Delta$ is a subset of the set composed of all the possible configurations, and the most relevant to evaluate learning.
Figure \ref{fig:bands} \textbf{a} shows the probability that, if we sample at T=1 from the Boltzmann distribution defined by the weights of the RBM, we get a configuration in $\Delta$. Figure  \ref{fig:bands} shows how such probability evolves as a function of the training epochs, for the forward, reverse, and classical cases respectively. It also shows the evolution of the probability associated with each element of the dataset, dividing the plot in $30$ distinct bands.

In the classical learning case, the probabilities appear more homogeneous than in the quantum cases. The training with forward--annealing runs at a slower pace but all the probabilities are growing steadily. Regarding the training with reverse--annealing, it is manifest that it resulted in an increased overall probability for configurations belonging to the set $\Delta$. In particular, the probability to extract a configuration in $\Delta$ from the RBM trained with reverse--annealing is approximately double with respect to the probability in the forward-trained case. Nonetheless, the average log-likelihood is similar in the two cases. The reason resides in the low probability associated with some configurations by the RBM trained with reverse-annealing. The black line in Figure \ref{fig:bands} \textbf{a} represents the summation of the probabilities of the $15$ less probable images. Since the $4\times 4$ BAS dataset is composed of $30$ images, the black line divides the dataset into two equal parts, not equally represented). The RBM trained with reverse--annealing is the best of the three cases if we look at the total probability of the dataset, at the cost of a lower probability of the lesser represented images. Such behavior could be connected to the fact that reverse annealing has a higher probability to produce configurations that are similar to those appearing in the dataset.

The different sampling distribution brings two competing consequences. At the beginning of the learning, the gradient estimation is influenced more by the configurations belonging to the dataset, which results in lowering their energy faster. This effect is exactly what the reverse annealing was intended for (Figure \ref{fig:bands} \textbf{a}). Later during learning, some configurations belonging to $\Delta$ are driven at low energies by the weights update and their probability is heavily underestimated.
The training algorithm does not capture the relevance that such configurations have in the partition function, and thus allows them to lower energy at each step. 

The sampling described above may be affected by the extraction being too bounded to the dataset elements (which could be overcome by optimizing the annealing schedule $s(t)$) or to a high sampling temperature, which in turn could be overcome by optimizing $\alpha$.

As a last consideration, we suggest evaluating an alternative and complementary figure of merit together with the average Log--Likelihood to score a model. A potential issue connected to Log--Likelihood manifests if some images have low probability. Indeed, they carry a lower  Log--Likelihood score, but it could not be informative about the capability to reconstruct data. Indeed, during reconstruction part of the visible units are clamped, thus many of the other configurations will have zero probability to be produced by the RBM. A better scoring method should only consider the ability of the final RBM to reconstruct each element of the training set. In our case, the RBMs have been trained on the whole dataset, so the training set and the test set coincide. 

Figure \ref{fig:results} \textbf{a} presents the reconstruction score as the probability for an RBM to reconstruct one of the four central pixels of a BAS image, averaged over the four pixels and each image in the set. The reconstruction is performed by keeping the outer pixels set to the correct values. The number of steps for Gibbs sampling is $n_g=500$. Note that the choice for the clamped units reduces dramatically the probability for configurations corresponding to different images in the dataset to take part in the reconstruction process since for each choice of the outer pixels there is a single dataset image compatible. Each colored band represents the probability that the corresponding image of the dataset is sampled from the Boltzmann distribution associated with the RBM.
The reconstruction score confirms the quality of the learning already evaluated by the average Log--Likelihood. As expected, the low--probability images in the training with reverse--annealing have a reduced impact on the score. At epoch $1000$, the reconstruction score of the reverse and forward method are statistically close, but reverse annealing exhibits better performances in previous epochs (Figure \ref{fig:results} \textbf{a}). 

In general, the reverse annealing schedule introduces a meaningful search method during the learning process of an RBM on a quantum computer. Despite its slower learning rate if compared to a classical machine in terms of epochs, one should remember that the total computational time is accounted for by the product of the number of epochs with the computational time per epoch. Therefore, the advantage of the adiabatic quantum computer to manage RMBs and more generally BMs resides on its ability to be employed once the conventional hardware fails as soon as the number of qubits of the quantum processor can handle the size of the problem. 
\section{Methods}
\label{sec:Methods}

\subsection{Embedding the problem}
To compare our results with existing literature, a $16\times16$ RBM is considered. Differently from the one unit--to--one qubit mapping constrained by the topology of the quantum circuit to return a sparse RBM because of the missing connections \cite{benedetti2016estimation}, we apply embedding technique to synthesize a full $16\times16$ RBM. The graph of the chosen embedding is shown in Supplementary Figure 3. In such an embedding, each unit of the RBM is mapped on four distinct physical qubits, which are bonded together by a strong negative coupling $J_C$. We used the strongest available coupling, $J_C=-1$,  in the dimensionless units used in the manual of the processor \cite{dwave2019technical}. The strong ferromagnetic coupling forces the four qubits to behave as a single two--state system, therefore behaving as a single virtual qubit. Indeed, each configuration where two of the four qubits are antiparallel is strongly energetically forbidden. This choice extends the connectivity of the hardware, allowing to implement all the $256$ desired connections. The embedding is reported in the Supplementary Note \textbf{3}. In principle, we could have used a different value for $J_C$. Supplementary Figure 4 shows that for $J_C=-0.5$ the learning gains no apparent advantage. We thus chose to use $J_C=-1$ throughout the work to minimize the probability of breaking the virtual qubits.

As the embedding requires $128$ qubits, we should in principle embed the problem in $16$ distinct positions in the superconductive chip, since the number of total qubits equals $2048$. Unfortunately, due to faulty qubits, the problem can be mapped in $8$ distinct positions. As a consequence, each annealing cycle produces $8$ samples. Running $8$ processes in parallel does not only sensibly reduce the computational cost per sample but is also useful to reduce the impact of sources of systematical errors that can be present in the AQC.

\subsection{Weight initialization and parameters of the learning process}

Weights and biases of the RBM were initialized by extracting random values from a Gaussian distribution with $\mu=0$, $\sigma=2$, truncated in $[-3,3]$. The learning rate was set to $\eta=0.15$. In the forward annealing case, an annealing time of $2 \mu$s was chosen. For the reverse annealing case, we set a reverse step of $1 \mu$s, followed by $18 \mu$s of pause and finally a last $1 \mu$s of forward annealing. During the pause $s(t)=0.2$.

For the classical case, training was performed with $n_g=200$. For the quantum case, we performed $150$ annealing cycles in each training epoch, both for the forward and reverse annealing. Considering that each annealing cycle outputs $8$ configurations, it sums up to $1200$ configurations.

\subsection{Temperature estimation}

For the annealing to produce configurations according to the Boltzmann distribution at $T=1$, one needs to properly rescale the parameters of the RBM before passing them to the AQC (Equation \ref{eq:mapping_parameters_RBM_to_AQC}). It is known that the best results are obtained by fine-tuning the parameter $\alpha$ by estimating the effective temperature $T_\text{eff}$ at any stage during the training \cite{benedetti2016estimation}. Nonetheless, the training can be effective even assuming that $T_\text{eff}$ is constant during training. In such a case, a wrong guess for $T_\text{eff}$ can lead to severe suboptimal learning scores.

We decided to avoid a dynamic estimation of $T_\text{eff}$ to simplify the experiment design, to test the robustness of the approach and to keep the focus on the differences between the forward annealing and the reverse annealing procedure. We initialized an RBM with random weights and biases extracted from the probability distribution defined in the previous section and trained it with forward annealing using several different values for the rescaling parameter $\alpha$. The value $\alpha=0.32$ resulted in the best score after $1000$ iterations. Note that such value is significantly different from others reported in the literature for the sparse RBM, as $\alpha\sim0.11$ in Ref.  \cite{benedetti2016estimation} that we confirmed in our experiment (Supplementary Figure 3). The difference is attributed to the use of virtual qubits composed of four physical qubits bounded by strong negative couplings. Indeed, higher values for the couplings between qubits can influence the freezing process and thus change $T_\text{eff}$ \cite{amin2015searching}, \cite{korenkevych2016benchmarking}. 

We trained an RBM with the sparse embedding of Ref. \cite{benedetti2017quantum}, which indeed resulted in $\alpha~0.11$ as the best value for the rescaling parameter. The corresponding learning curve is represented in Supplementary Figure 2. We evaluated the connected RBM with initial weights and biases extracted from a gaussian with $\mu=0$, $\sigma=1.5$, truncated in $I_W=[-2,2]$ (i.e. weights and biases are smaller than those used to produce the plots in Figure \ref{sec:results} \textbf{a} and \textbf{b}). In such a case, the best value for $\alpha$ among those tested is $\alpha=0.33$. The corresponding learning curve is represented in Supplementary Figure 1.

Note that the currents inside the device possess a maximum and minimum programmable value. Thus, it is possible that during learning the weights to be passed to the AQC exceed those boundaries. In our implementation, weights and biases have never reached values close to the boundaries.

\section{Conclusions}
\label{sec:conclusions}

    Boltzmann Machines can be trained with an algorithm that exploits AQCs with the spirit of achieving a computational advantage over the classical method. We showed that the use of embedding techniques does preserve the quality of produced configurations, and thus it results in significantly better scores thanks to the increased connectivity.
    As opposed to the usual algorithm based on forward annealing schedule, we implemented a semantic quantum search based on reverse annealing schedule. We showed that such an algorithm quickly raises the sampling probability of a subset of the configurations set corresponding to elements of the dataset and can achieve good reconstruction scores in slightly less training epochs than forward annealing. 
    Reverse annealing captures the benefit of starting the algorithm by exploiting the full information provided by the elements of the dataset. It leads to a sampling probability of elements of the dataset which is double that of the forward annealing and higher than that of the classical method. 
    
    Our results, combined with hyperparameter optimization of the annealing schedule and temperature estimation techniques pave the way towards the exploitation of both restricted and unrestricted Boltzmann machines as soon as new generations of hardware with increased connectivity will be available.
	
\section{Acknowledgements}
	The EP gratefully acknowledges D--Wave for having partially granted the access to the D--Wave 2000Q\textsuperscript{TM} System processor, used to obtain the experimental results presented in this article. EP and LR would like to thank Daniele Ottaviani and Carlo Cavazzoni of CINECA for useful discussions. 

\section{Data availability}
	The data that support the findings of this study are available from the corresponding Author upon reasonable request.
	
\section{Competing interest}
	The Authors declare that there are no competing interests
	
\section{Author contributions}
	 LR developed the quantum algorithm and analysed the data, CD contributed to theoretical aspects of Gibbs sampling and quantum dynamics, EP conceived and coordinated the research. All the Authors discussed the results and contributed to the writing of the manuscript.

\bibliography{biblio}

 \end{document}